\documentclass{iaus}
\usepackage{graphicx}

\title[Complex Envelopes] 
{Morphological Complexity of Protostellar Envelopes}

\author[Tobin et al.]   
{John J. Tobin$^1$, Lee Hartmann$^1$, Edwin Bergin$^1$, Leslie W. Looney$^2$, Hsin-Fang Chiang$^2$,
Fabian Heitsch$^3$}

\affiliation{$^1$Department of Astronomy, University of Michigan, Ann
Arbor, MI 48109; jjtobin@umich.edu\\[\affilskip]
$^2$Department of Astronomy, University of Illinois at
Champaign/Urbana, Urbana, IL 61801\\[\affilskip]
$^3$Department of Physics and Astronomy, University of North Carolina-Chapel Hill, Chapel Hill, NC}

\pubyear{2010}
\volume{270}  
\pagerange{1-4}
\jname{Computational Star Formation}
\editors{Alves, Elmegreen, Girart, \& Trimble, eds.}
\begin{document}

\maketitle

\begin{abstract}
Extinction maps at 8$\mu$m from the Spitzer Space Telescope
show that many Class 0 protostars exhibit
complex, irregular, and non-axisymmetric structure within the 
densest regions of their dusty envelopes.
Many of the systems have highly irregular and non-axisymmetric
morphologies on scales $\sim$1000 AU, with a quarter of the
sample exhibiting filamentary or flattened dense structures. 
Complex envelope structure is observed in regions spatially distinct
from outflow cavities, and the densest structures often show
no systematic alignment perpendicular to the cavities.
We suggest that the observed envelope complexity
is the result of collapse from protostellar
cores with initially non-equilibrium structures. 
The striking non-axisymmetry in many envelopes could provide favorable
conditions for the formation of binary systems. 
We then show that the kinematics around L1165 as probed with N$_2$H$^+$ 
are indicative of asymmetric infall; the velocity gradient is not perpendicular to the
outflow.


\end{abstract}

\firstsection 
\section{Introduction}
 Sphericity and axisymmetry have been standard assumptions on which
our theoretical understanding of star formation has rested for some time
\cite[(e.g. Shu 1977; Terebey et al. 1984; Galli \& Shu 1993;
 Hartmann et al. 1996)]{shu1977,tsc1984,galli1993,hartmann1996}.
However, it is not clear whether or not envelopes around protostars
are accurately described by symmetric models. The shapes of dense cores
have been studied on large scales ($>$0.1 pc) 
using molecular line tracers of dense gas
\cite[(Benson \& Myers 1989; Myers et al. 1991)]{benson1989,myers1991}. While the ammonia
cores appeared compact and round due to low resolution, associated IRAS sources
are often located off-center from the line emission peaks indicating non-axisymmetry.
Furthermore, emission from the millimeter-line tracers appears
quite complex on scales outside the ammonia emission\cite[(Myers et al. 1991)]{myers1991}.

Recently, observations with the \textit{Spitzer Space Telescope} have given a high-resolution
view of envelope structure in extinction at 8$\mu$m against Galactic
background emission \cite[(Looney et al. 2007; Tobin et al. 2010)]{looney2007,tobin2010}.
This method enables us to observe the structure of collapsing protostellar envelopes
on scales from $\sim$1000AU to 0.1 pc for the first time with a mass-weighted tracer. In contrast,
single-dish studies of envelopes using dust emission in the sub/millimeter
regime generally have lower spatial resolution. The continuum
emission depends upon temperature and density, while
molecular tracers are additionally affected by complex chemistry.

In this contribution, we present IRAC 8$\mu$m extinction maps of envelopes around
 Class 0 protostars.
Most of the envelopes in our sample are found to be irregular and non-axisymmetric.
We suggest that these envelopes may be important for binary formation as collapse
models have shown that asymmetric structure can induce small-scale fragmentation of the
infalling envelope. We also show initial results in analyzing the kinematic structure
of these envelopes using the dense gas tracers N$_2$H$^+$.

\begin{figure}[ht]
 \vspace*{-0.5 cm}
\begin{center}
 \includegraphics[scale=0.6, angle=-90]{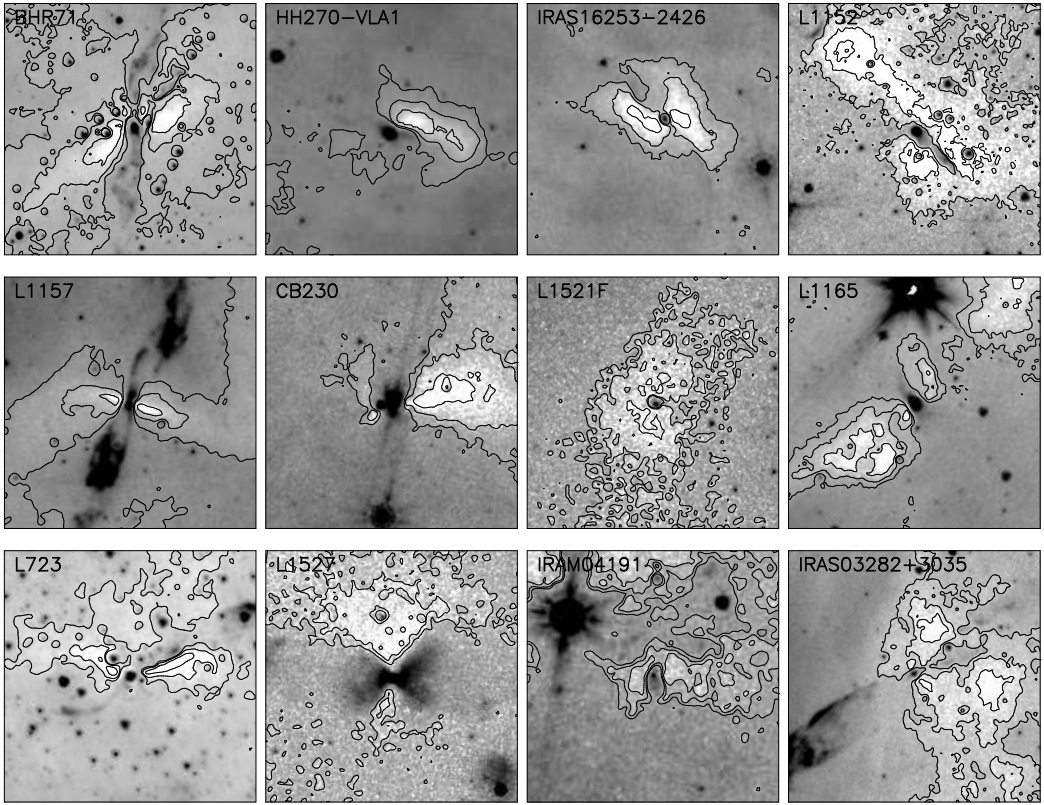} 
 \caption{IRAC 8$\mu$m images with 8$\mu$m optical depth contours overlaid for a selection
of the protostars from \cite[Tobin et al. (2010)]{tobin2010}. From left to right the columns
are filamentary envelopes, one-sided envelopes, quasi-symmetric envelopes, and irregular envelopes
respectively.}
\end{center}
\vspace*{-0.75 cm}
\end{figure}

\section{Asymmetric Envelopes in Extinction}
In Figure 1, we display a selection of 12 systems (out of 22 from Tobin et al. (2010)) for which we have detected
an envelope in extinction. Each panel shows the 8$\mu$m image of the protostar with
optical depth contours overlaid. In most images, the outermost extinction contour
corresponds to  A$_V$ $\sim$5-10 and going up to A$_V \sim 30$.

The most striking feature of the 8$\mu$m extinction maps is the irregularity of 
envelopes in the sample. Most envelopes show high degrees of non-axisymmetry;
in most cases, spheroids would be an inaccurate representation of the
structure. Some of the most extreme examples have most extincting material mostly on
one side of the protostar (e.g. CB230, HH270 VLA1) or the densest structures are
curved near the protostar (e.g. BHR71, L723). The structures seen in extinction at 8$\mu$m
do not seem to be greatly influenced by the outflow. \cite[Tobin et al. (2010)]{tobin2010}
demonstrated that the outflow cavities of these sources are generally quite narrow and 
the dense material detected in extinction is often far from the outflow cavities and thus
unlikely to be produced by outflow effects.

For convenience, we categorize the systems into 4 groups according to their morphology, 
though some systems have characteristics of multiple groups.
The first column of Figure 1 shows the envelopes that have a highly
 filamentary/flattened morphologies; column 2 shows envelopes that
 have most material on one side of the protostar; column 3 shows the protostars
whose envelopes are more or less spheroidal in projection;
column 4 shows envelopes that do not strictly fall within the above categories.

\section{Kinematic Structure}
In addition to morphological structure, the kinematic structure of the envelope resulting
from the non-axisymmetric envelope is of great importance for the subsequent evolution of the 
system. Figure 2 shows the N$_2$H$^+$ ($J=1\rightarrow 0)$ observations for the protostar L1165 from the CARMA
millimeter array. The N$_2$H$^+$ emission shows that the envelope is elongated along the 
larger filament consistent with the IRAC 8$\mu$m image and the peak emission is offset from
the protostar; the envelope also remains filamentary down to small scales.
The centroid velocity map shows a strong velocity gradient across
the envelope, 0.35 km/s over 2400 AU; though, the linewidth indicates that the gradient could be larger.
At present, it is difficult to differentiate between \textit{projected}
infall or rotation giving rise to the velocity structure; in either case the infall motions will be 
complex. Given that the N$_2$H$^+$ velocity gradient is $\sim$30$^{\circ}$
from perpendicular to the outflow and the HCO$^+$ gradient is perpendicular it seems possible
that the angular momentum vector may change with time.

\begin{figure}[ht]
\vspace*{0.0 cm}
\begin{center}
 \includegraphics[scale=0.55,angle=-90]{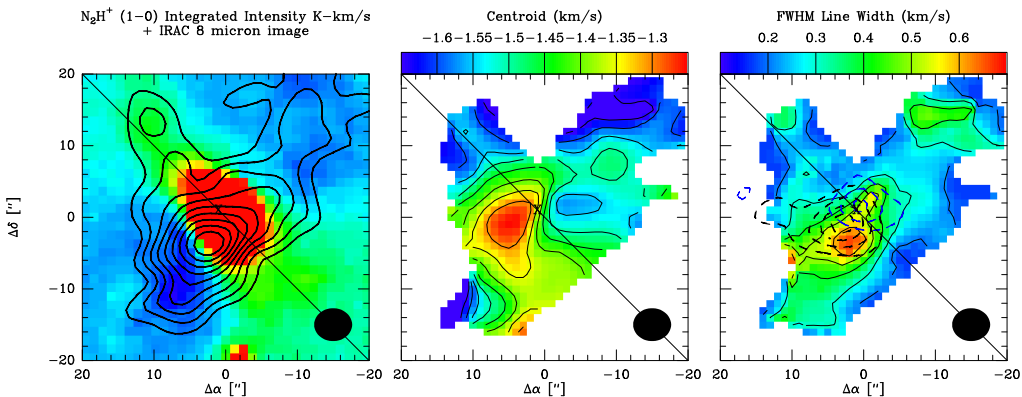} 

 \caption{N$_2$H$^+$ ($J=1\rightarrow 0)$ observations for the protostar L1165; in all panels
the protostar is marked with an X and the outflow axis is marked with a straight
line. Left: IRAC 8$\mu$m image (\textit{color scale}) with
N$_2$H$^+$ ($J=1\rightarrow 0)$ integrated intensity (\textit{contours}). 
Middle: Centroid velocity of N$_2$H$^+$ emission 
derived from fitting the hyperfine lines; notice the sharp gradient across the protostar 
which is not perpendicular to the outflow. Right: Full-width half maximum of N$_2$H$^+$ lines;
note that the linewidth is greatest at the peak of N$_2$H$^+$ emission. The dashed contours in the
rightmost panel are the HCO$^+$ blue (\textit{blue}) and redshifted (\textit{black}) emission.}
\end{center}
\vspace*{-0.75 cm}
\end{figure}

\section{Implications of Non-Axisymmetric Collapse}

The envelope asymmetries may result from the initial cloud structure.
\cite[Stutz et al. (2009)]{stutz2009} recently surveyed pre-stellar/star-less cores using 8 and 24$\mu$m
extinction; their results, and those of \cite[Bacmann et al. (2000)]{bacmann2000}, showed that even
pre-collapse cloud cores already exhibit asymmetry. Given the initial asymmetries,
the densest, small-scale regions are likely to become more anisotropic during
 gravitational collapse \cite[(Lin et al. 1965)]{lin1965}.

\subsection{Binary Formation}

The smallest scales we observe, $\sim$1000 AU, is where angular momentum will begin
to be important as the material falls further in onto the disk; this is the scale
we probe with the N$_2$H$^+$ observations of L1165.
The envelope asymmetries down to small scales imply that infall
to the disk will be uneven; therefore,  non-axisymmetric infall
may play a significant role in disk evolution and the formation of binary systems. 
Several theoretical investigations \cite[(e.g. Burkert \& Bodenheimer 1993)]{burkert1993} 
showed that collapse of a cloud with just a small azimuthal perturbation can
form binary or multiple systems; thus, {\em large} non-axisymmetric perturbations
should make fragmentation even easier. Fragmentation can even begin before
global collapse in a filamentary structure \cite[(Bonnell et al. 1993)]{bonnell1993}.
Numerical simulations of disks with infalling envelopes 
\cite[(e.g. Kratter et al. 2009)]{kratter2009}
informed by the results of this study could reveal a more complete understanding
of how non-axisymmetric infall affects the disk and infall process.
Asymmetric infall may be taking place in at least L1165. However,
it is not known if this system is a binary because the requisite high-resolution
observations have not been taken. 

\subsection{Initial Conditions}

The envelopes we have shown are not obviously consistent with quasi-static, slow evolution,
which might be expected to produce simpler structures as irregularities have
time to become damped; one needs initial non-axisymmetric
structure to get strong non-axisymmetric structure later on. This raises the question
of the role of magnetic fields in controlling cloud dynamics. In some models
\cite[(e.g. Tassis et al. 2007)]{tassis2007}, pre-stellar
cores would probably live long enough to adjust to more regular configurations; in addition,
collapse would be preferentially along the magnetic field, which would 
also provide the preferential direction of the rotation axis and therefore
for the (presumably magnetocentrifugally
accelerated) jets \cite[(Basu \& Mouschovias 1994)]{basu1994}. The
complex structure and frequent misalignment between collapsed structures
and outflows pose challenges for such a picture. Alternatively, this could indicate that the 
topology of magnetic field is complex and not initially well ordered.

In contrast, more recent numerical simulations
\cite[(e.g. see review by Ballesteros-Paredes et al. 2007; Offner \& Krumholz 2009)]{ballesteros2007,offner2009}
suggest that cores are the result
of turbulent fluctuations which naturally produce more complex structure
with less control by magnetic fields amplified by subsequent gravitational contraction and collapse.
Thus, the structure of protostellar envelopes indicates that the dynamic, turbulent
model of rapid star formation seems more correct. However, recent work has indicated that
turbulence can reduce the time needed for ambipolar diffusion to take place which may allow
complex structures to form \cite[(Basu \& Ciolek 2009)]{basu2009}.

\end{document}